\documentclass[aps,pra,reprint,superscriptaddress,nofootinbib]{revtex4-2}
\usepackage{amsmath,amssymb,bm}
\usepackage{physics}
\usepackage{graphicx}
\usepackage{appendix}

\begin{document}

\title{Multi-slit time-reversed Young interference: source-space grating laws, quadratic-phase effects, and Talbot-like revivals}

\author{Jianming Wen}
\email{jwen7@binghamton.edu}
\affiliation{Department of Electrical and Computer Engineering, Binghamton University, Binghamton, New York 13902, USA}

\begin{abstract}
We develop a compact theory of time-reversed Young (TRY) interference beyond the symmetric two-slit geometry by considering equally spaced three-slit, finite $N$-slit, and infinite periodic slit arrays. In the TRY configuration, a point emitter illuminates the aperture, a position-fixed detector records the signal, and the response is reconstructed in source space by correlating the detector record with the source-coordinate label. We show that the three-slit case already reveals the essential new physics beyond two slits: a quadratic Fresnel phase survives, modifies the reconstructed interference law, and lifts the nominal dark fringes in the generic case. For a general equally spaced $N$-slit array, we identify the exact reconstructed response and show that the familiar textbook grating factor is recovered only when the quadratic phase is negligible, compensated, or reduced to a common phase across the array. In that ideal limit, the reconstructed peaks are source-space analogues of classical grating orders rather than outgoing diffraction beams. For an infinite periodic TRY array, we further show that the same discrete quadratic phase generates full and fractional Talbot-like revivals in source space, governed by a reciprocal-distance condition rather than the conventional Talbot propagation law. These results show that the symmetric two-slit TRY geometry is exceptional, while multi-slit TRY systems naturally combine source-space discrimination with sensitivity to aperture-wide phase structure and periodic-array revival physics.
\end{abstract}

\maketitle

\section{Introduction}

Young's double-slit experiment is one of the canonical paradigms of optical interference~\cite{Young1804,BornWolf,GoodmanFO,Hecht}. In its textbook form, a source illuminates an aperture, the field propagates downstream to an observation screen, and the interference pattern is read out as a function of angle or transverse screen position. The time-reversed Young (TRY) configuration~\cite{Wen2025} reverses this logic in a nontrivial way: a laterally addressable point emitter replaces the conventional source, a position-fixed detector replaces the observation screen. The interference law is then reconstructed as a function of source coordinate rather than observation coordinate. The pattern is therefore not recorded directly across a screen, but assembled from the fixed-detector signal after sorting the data by source position. Operationally, this reconstructed signal is not an ordinary first-order spatial interference pattern. Rather, it is a hybrid second-order correlation formed by correlating the detector output with the source-coordinate label~\cite{WenHybrid2026}.

The original TRY proposal~\cite{Wen2025} has already led to several developments that sharpen both its physical interpretation and its practical relevance. A recent hybrid-coherence formulation placed the reconstructed TRY signal on an explicit second-order footing by combining the detector intensity with a source-coordinate projector, thereby embedding the effect more naturally in a quantum-optical framework \cite{WenHybrid2026}. Two subsequent works further showed that the source plane in TRY is not merely a scan coordinate, but can also serve as a programmable measurement basis. One introduced differential source-basis encoding for superresolved local parameter estimation \cite{WenDiff2026}; the other established optimal null-constrained source-basis sensing and quantified the information cost associated with imposing a metrological null \cite{WenNull2026}. Taken together, these results show that TRY is not only a conceptual reversal of Young's experiment, but also a flexible fixed-detector architecture for interferometric sensing.

The symmetric double-slit TRY geometry is especially simple because the quadratic Fresnel phase associated with the slit coordinates identical for the two paths and therefore cancels out of the response. The natural next question is whether this simplification survives for three slits and, more generally, for an $N$-slit grating. It does not. For three or more slits, the slit-position-squared phase is no longer common to all paths, and the exact TRY response acquires a quadratic phase structure that is absent in the symmetric two-slit case and has no analogue in the ideal textbook Fraunhofer grating law~\cite{GoodmanFO,Hecht,LoewenPopov}. The central theoretical issue is therefore not merely how to extend the two-slit formula to more paths, but how the surviving quadratic phase reshapes the reconstructed interference law in source space.

That quadratic structure also suggests a broader connection with periodic diffraction. In conventional near-field optics, the Talbot effect arises from a quadratic Fresnel phase acting on a periodic grating and gives rise to full and fractional self-imaging \cite{Talbot1836,Rayleigh1881,Patorski1989,BerryKlein1996,WenTalbot2013,Testorf2010,Mansuripur2011}. The present TRY problem is different in both geometry and observable: the relevant quantity is a source-labeled fixed-detector response rather than a propagated transverse intensity distribution. Nevertheless, the same discrete quadratic-phase mechanism suggests that periodic TRY arrays may exhibit Talbot-like revivals in source space. Clarifying this analogy, while also making precise where TRY departs from traditional Talbot physics, is one of the main aims of the present work.

This work has three main objectives. First, we develop a compact theory of TRY interference for equally spaced three-slit and finite $N$-slit arrays. Second, we identify the distinct roles of the linear, source-controlled phase and the quadratic Fresnel phase, and show why the latter becomes directly observable only beyond the two-slit case. Third, we show that, for an infinite periodic TRY array, the same discrete quadratic phase generates full and fractional source-space revivals that are strongly analogous to Talbot physics, yet remain distinct from conventional transverse-field self-imaging. The resulting picture is that multi-slit TRY unifies source-space grating behavior, fixed-detector operation, and aperture-wide phase sensitivity within a single framework.

\section{General TRY formulation}
As schematically shown in Fig.~\ref{fig:nslit_scheme}(a), we place the slit plane at $z=0$ and use $x$ as the transverse coordinate. The slit array is centered at $x=0$, so the slit coordinates are distributed symmetrically about the origin. A point emitter is located at transverse coordinate $x_s$ in the source plane $z=-z_1$, where $z_1>0$ is the source-to-aperture distance. A point detector is fixed at transverse coordinate $x_D$ in the detection plane $z=+z_2$, where $z_2>0$ is the aperture-to-detector distance. The optical field is monochromatic, with wavelength $\lambda$ and wavenumber $k=2\pi/\lambda$.

\begin{figure}[t]
\includegraphics[width=0.9\columnwidth]{figure1.png}
\caption{Time-reversed Young (TRY) geometries for an equally spaced $N$-slit array and for the three-slit case. A point emitter at transverse coordinate $x_s$ illuminates the slit plane at $z=0$, and a point detector is fixed at transverse coordinate $x_D$ in the detector plane. The source-to-slit and slit-to-detector distances are $z_1$ and $z_2$, respectively; in panel (a), the slit coordinates are $x_n=p_nd$ with spacing $d$, while panel (b) shows the special three-slit geometry.}
\label{fig:nslit_scheme}
\end{figure}

We consider an equally spaced $N$-slit array with slit spacing $d$. The slit-center coordinates are written as
\begin{equation}
x_n=p_nd,\quad p_n=n-\frac{N-1}{2},\quad n=0,1,\dots,N-1,\label{eq:pn}
\end{equation}
where $p_n$ is the centered slit index. This notion makes the array symmetry explicit: for every slit at $+x$, there is a corresponding slit at $-x$. Accordingly, $p_n$ is integer for odd $N$ and half-integer for even $N$. The exact source-to-slit and slit-to-detector path lengths are
\begin{equation}
r_n=\sqrt{z_1^2+(x_s-x_n)^2},\;R_n=\sqrt{z_2^2+(x_D-x_n)^2},\label{eq:path_exact}
\end{equation}
so that the total optical path through slit $n$ is
\begin{equation}
L_n=r_n+R_n.
\end{equation}
Throughout this work, we adopt a scalar, monochromatic, narrow-slit, paraxial model. The paraxial conditions require
\begin{equation}
|x_s-x_n|\ll z_1,\quad|x_D-x_n|\ll z_2\label{eq:paraxial_cond}
\end{equation}
for all slits. The narrow-slit approximation allows each slit to be represented by a single complex amplitude, and we further neglect slit-to-slit amplitude variation across the array. Under these assumptions, the path lengths~\eqref{eq:path_exact} may be Taylor expanded as
\begin{equation}
r_n\approx z_1+\frac{(x_s-x_n)^2}{2z_1},\quad R_n\approx z_2+\frac{(x_D-x_n)^2}{2z_2}.
\label{eq:paraxial_paths}
\end{equation}

Substituting Eq.~(\ref{eq:pn}) into Eq.~(\ref{eq:paraxial_paths}), and then combining the result in Eq.~(3) gives
\begin{equation}
kL_n=\Phi_c+\beta p_n^2-\gamma p_n,
\label{eq:phase_form}
\end{equation}
with
\begin{align}
\Phi_c&=k\left(z_1+z_2+\frac{x_s^2}{2z_1}+\frac{x_D^2}{2z_2}\right),\label{eq:Phi}\\
\beta&=\frac{kd^2}{2}\left(\frac{1}{z_1}+\frac{1}{z_2}\right),\label{eq:beta}\\
\gamma&=kd\left(\frac{x_s}{z_1}+\frac{x_D}{z_2}\right).\label{eq:gamma}
\end{align}
Here $\Phi_c$ is a global phase and therefore does not affect the intensity. The parameter $\gamma$ is the linear inter-slit phase, while $\beta$ is the quadratic Fresnel phase associated with the squared slit coordinates. Physically, $\gamma$ governs the ordinary array oscillation, whereas $\beta$ measures how strongly the finite propagation geometry imprints aperture-wide curvature across the array.

It is useful to express $\gamma$ in geometric form. In the paraxial regime, we define the source-side and detector-side opening angles by
\begin{equation}
\tan\theta_s=\frac{x_s}{z_1}\approx\theta_s,\quad
\tan\theta_D=\frac{x_D}{z_2}\approx\theta_D,\label{eq:angles}
\end{equation}
so that Eq.~(\ref{eq:gamma}) becomes
\begin{equation}
\gamma\approx kd(\theta_s+\theta_D),\label{eq:gamma_angles}
\end{equation}
Thus, $\gamma$ is the net adjacent-slit phase induced by the source-side and detector-side geometry. In particular, for a centered detector,
\begin{equation}
x_D=0\quad\Rightarrow\quad\gamma\approx kd\,\theta_s.\label{eq:gamma_sourceangle}
\end{equation}
This source-side tunability is one of the main practical features of TRY: the reconstructed response can be controlled entirely from the source side while the detector remains fixed. In this sense, a TRY array can function as a single-pixel analyzer of source-side opening angle or source position. The same fixed-detector architecture is especially attractive in detector-limited platforms where array detectors are costly, noisy, slow, or unavailable~\cite{Edgar2019,Gibson2020,Chan2008,Wang2023}. It also makes the source plane a natural programmable space, consistent with recent source-basis sensing developments in TRY~\cite{WenDiff2026,WenNull2026}.

Let $A$ denote the common complex amplitude transmitted by each slit, and define $I_0=|A|^2$ as the corresponding single-slit intensity scale. The total field at the fixed detector is then
\begin{equation}
E_N(x_s;x_D)=Ae^{i\Phi_c}\sum_{n=0}^{N-1}e^{i\beta p_n^2-i\gamma p_n},\label{eq:EN}
\end{equation}
and the reconstructed TRY response is
\begin{equation}
I_{\mathrm{TRY}}^{(N)}(x_s)=|E_N(x_s;x_D)|^2=I_0\left|\sum_{n=0}^{N-1} e^{i\beta p_n^2-i\gamma p_n}\right|^2.\label{eq:ITRYN}
\end{equation}
The dependence on $x_s$ should be understood operationally: the pattern is reconstructed by correlating the fixed-detector record with the source-coordinate label, so the multi-slit TRY response retains the same hybrid second-order character as the original two-slit scheme. For the same reason, Eq.~(\ref{eq:ITRYN}) is not a conventional observation-plane first-order intensity profile and therefore does not, in general, factor into a classical single-slit envelope multiplying an array factor.

A complementary form follows by expanding the modulus square:
\begin{equation}
I_{\mathrm{TRY}}^{(N)}(x_s)=NI_0+2I_0\sum_{0\le m<n\le N-1}
\cos\!\left[\beta(p_n^2-p_m^2)-\gamma(p_n-p_m)\right].\label{eq:pairwise}
\end{equation}
The first term is the incoherent sum of the $N$ single-slit contributions, whereas the second term contains the coherent pairwise interference. The key point is that the phase depends not only on the pair separation $p_n-p_m$, but also on the pair center through
\[
p_n^2-p_m^2=(p_n-p_m)(p_n+p_m).
\]
Thus, equal-separation pairs are not generally phase matched. This is precisely why the exact TRY grating does not automatically reduce to the textbook array factor.

The quadratic phase $\beta$ deserves special emphasis because it is the structural feature that distinguishes multi-slit TRY from both the symmetric two-slit case and the ideal textbook grating law. Its role is not simply to add an overall phase; rather, it redistributes the relative phases of different slit pairs across the aperture. For a fixed pair separation, the additional factor $\beta(p_n^2-p_m^2)$ depends on the pair center through $p_n+p_m$, so slit pairs with the same spacing do not, in general, interfere in phase. This has several immediate consequences. First, the exact reconstructed pattern is generally deformed away from the ideal array factor. Second, the effect becomes stronger as the array width increases, since the largest values of $p_n^2$ scale as $[(N-1)/2]^2$. Third, in the three-slit case, it lifts the nominal dark fringes, as shown explicitly in Sec.~III. In physical terms, $\beta$ quantifies how strongly the finite propagation geometry imprints aperture-wide curvature across the array. This is why the multi-slit TRY response becomes naturally sensitive to defocus, wavefront curvature, slit-position errors, and array-scale phase defects or nonuniformity \cite{Roddier1988,Paxman1992,Chanan1999,PlattShack2001,Primot2003,Tyson2015}. Such sensitivity also makes the fixed-detector TRY architecture a natural candidate for aperture self-calibration and compact wavefront diagnostics. The same formalism further clarifies why odd and even arrays are not fully equivalent: integer-centered and half-integer-centered index sets organize the quadratic phase differently across the aperture, providing an additional structural degree of freedom for engineering the exact source-space response.

\section{Three-slit TRY law}
We now specialize the general formulation of Sec.~II to three equally spaced slits centered at $x=0$, with slit coordinates
\begin{equation}
x_{-1}=-d,\quad x_0=0,\quad x_{+1}=+d,\label{eq:3slit_positions}
\end{equation}
where $d$ is the adjacent slit spacing. The corresponding geometry is shown in Fig.~1(b). Relative to the two-slit case, the essential new feature is that, under the quadratic Fresnel phase, the middle slit at $x=0$ is no longer equivalent to the two outer slits.

Using Eq.~(\ref{eq:phase_form}), the phases accumulated through the three slits are
\begin{equation}
kL_{-1}=\Phi_c+\beta+\gamma,\;kL_0=\Phi_c,\;kL_{+1}=\Phi_c+\beta-\gamma,
\end{equation}
so the total field at the fixed detector, from Eq.~\eqref{eq:EN}, becomes
\begin{align}
E^{(3)}(x_s;x_D)&=Ae^{i\Phi_c}\left[e^{i(\beta+\gamma)}+1+e^{i(\beta-\gamma)}\right]\nonumber\\
&=Ae^{i\Phi_c}\left(1+2e^{i\beta}\cos\gamma\right).\label{eq:E3b}
\end{align}
The reconstructed three-slit TRY intensity is therefore
\begin{align}
I_{\mathrm{TRY}}^{(3)}(x_s)&=I_0\left|1+2e^{i\beta}\cos\gamma\right|^2\nonumber\\
&=I_0\!\left(1+4\cos\beta\cos\gamma+4\cos^2\gamma\right)\label{eq:I3}\\
&=I_0\!\left(3+4\cos\beta\cos\gamma+2\cos 2\gamma\right).\nonumber
\end{align}
Equation~(\ref{eq:I3}) is the exact three-slit TRY law within the present paraxial narrow-slit model. Its structure is physically transparent. The term $4\cos^2\gamma$ is the familiar three-path oscillation governed by the linear phase $\gamma$. The additional term $4\cos\beta\cos\gamma$ is the genuinely beyond-two-slit correction produced by the quadratic phase $\beta$. The reason is simple: the two outer slits share the same squared coordinate, $x^2=d^2$, whereas the middle slit has $x^2=0$. Consequently, the outer--outer interference is insensitive to $\beta$, while each outer--middle interference term retains it. The middle slit is therefore distinguished not by its spacing, which remains symmetric, but by its different squared transverse coordinate.

This distinction has an immediate observable consequence. The extrema of Eq.~(\ref{eq:I3}) satisfy
\begin{equation}
\sin\gamma=0\quad \text{or} \quad\cos\gamma=-\frac{\cos\beta}{2},\label{eq:extrema}
\end{equation}
and, at the second set of extrema, the minimum intensity is
\begin{equation}
I_{\min}=I_0\sin^2\beta.\label{eq:darkfloor}
\end{equation}
Thus, the nominal dark fringes are generally lifted. Exact zeros occur only when the quadratic phase reduces to a common phase, for example when $\beta=0\pmod{2\pi}$. This is the central physical lesson of the three-slit TRY problem: once the aperture contains a central slit and an outer pair, Fresnel-phase nonuniformity becomes directly visible in the reconstructed source-space response. 

\begin{figure}[htpb]
\includegraphics[width=0.88\columnwidth]{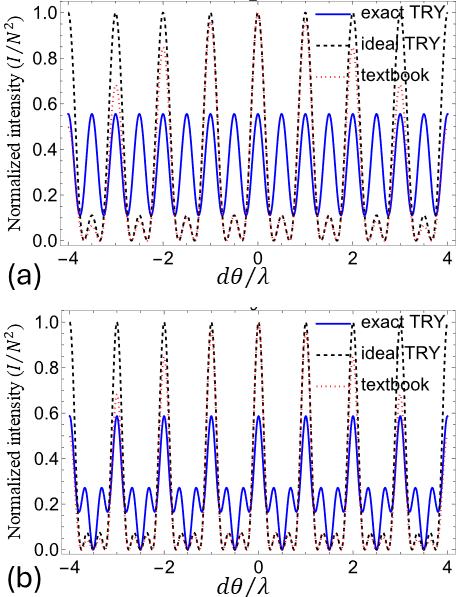}
\caption{Comparison of the exact TRY response (blue solid), the ideal TRY grating law (black dashed), and the textbook Fraunhofer result with single-slit envelope (red dotted) for (a) $N=3$ and (b) $N=4$. The horizontal axis is the normalized coordinate $d\theta/\lambda$, interpreted as the source-side opening-angle variable in TRY and as the observation angle in the textbook case. The simulations use a centered detector $x_D=0$, the exact TRY law $I^{(N)}_{\rm TRY}=\left|\sum^{N-1}_{n=0}e^{i\beta p^2_n-i2\pi(d\theta/\lambda)p_n}\right|^2$, the quadratic-phase choice $\beta p^2_{\rm max}=\pi/2$ with $p_{\rm max}=(N-1)/2$, and slit-width ratio $a/d=0.35$ for the textbook curve.}
\label{fig:comp_smallN}
\end{figure}

Figure~2(a) makes this mechanism visually explicit. The exact three-slit TRY curve is not just a displaced version of the ideal law \(I_{\mathrm{TRY,ideal}}^{(3)}\). Rather, the quadratic phase reshapes the pattern by lifting the minima and redistributing weight between the principal maxima and the nearby sidelobes. By contrast, the textbook three-slit curve~\cite{BornWolf,GoodmanFO,Hecht} is modified mainly by the single-slit envelope and therefore retains the usual Fraunhofer interpretation in observation angle. The figure therefore does more than provide a comparison plot: it cleanly separates two physically distinct nonideal effects. In TRY, the leading departure from the ideal array law is the quadratic Fresnel phase \(\beta\), whereas in the textbook case it is the observation-plane diffraction envelope \(I_1(\theta)\).

Equation~(\ref{eq:darkfloor}) also points naturally to a null-based sensing mode. In a nominally balanced three-slit configuration, the dark-fringe floor becomes a null observable for the quadratic phase. Small changes in propagation geometry, weak defocus, or slight wavefront-curvature perturbations are then converted directly into measurable fixed-detector signals \cite{Bracewell1978,SerabynColavita2001,OllivierMariotti1997,Mennesson2002,Lay2004,MartinacheIreland2018,WenNull2026}. In this sense, the three-slit TRY geometry becomes a natural fixed-port platform for null-based alignment sensing and weak-phase metrology, with the source-side control setting the operating point and the dark-floor level revealing departures from the compensated condition.

For a centered detector, Eq.~(\ref{eq:gamma_sourceangle}) gives $\gamma\approx kd\,\theta_s$, with $\theta_s$ being the source-side opening angle. Thus, the three-slit TRY response can be swept directly by source-side geometry while the detector remains fixed. This source-controlled tunability, together with the quadratic-phase-sensitive dark floor in Eq.~(\ref{eq:darkfloor}), is what makes the three-slit case the first genuinely new and practically useful extension beyond the symmetric two-slit TRY geometry.

\section{$N$-slit law and ideal TRY grating limit}
We now return to the general equally spaced $N$-slit array. The exact reconstructed TRY response has already been obtained in Eq.~(\ref{eq:ITRYN}), so the main question here is not how to rederive the array law, but under what conditions it reduces to a grating-like form and what that reduction means physically.

The first important point is that the two-slit case is exceptional. For $N=2$, the centered slit indices are $p_0=-1/2$ and $p_1=+1/2$, so that
\[
p_0^2=p_1^2=\frac14.
\]
The quadratic phase is therefore common to both slits and drops out of the response. For $N\ge 3$, however, the set $\{p_n^2\}$ is no longer uniform, so the quadratic phase survives and modifies the reconstructed pattern. The transition from two slits to three or more slits is therefore the point at which aperture-wide Fresnel structure becomes directly observable in the TRY response.

The ideal TRY grating law is recovered only when the quadratic phase factor $e^{i\beta p_n^2}$ is effectively common across the array. This occurs in three practically relevant situations. First, the total quadratic phase spread across the aperture may be small:
\begin{equation}
|\beta|\left(\frac{N-1}{2}\right)^2\ll 1.\label{eq:smallbeta}
\end{equation}
Second, the quadratic phase may be compensated by additional optical elements or by engineered slit-dependent phases. Third, if
\begin{equation}
\beta=2\pi\ell,\quad\ell\in\mathbb{Z},\label{eq:revival}
\end{equation}
then $e^{i\beta p_n^2}$ becomes a common phase for both odd and even $N$ and therefore does not affect the intensity. Equation~(\ref{eq:smallbeta}) also shows the main large-$N$ constraint: the compensated regime becomes progressively harder to maintain as $N$ increases, because the quadratic-phase sensitivity grows with the square of the array half-width. In all three cases, the array behaves as though only the linear phase remains operative.

Under these conditions, Eq.~(\ref{eq:ITRYN}) reduces to
\begin{equation}
I_{\mathrm{TRY,ideal}}^{(N)}(x_s)=I_0\left|\sum_{n=0}^{N-1} e^{-i\gamma p_n}\right|^2.
\label{eq:IN_ideal_start}
\end{equation}
Using $p_n=n-(N-1)/2$, we write
\begin{equation*}
\sum_{n=0}^{N-1}e^{-i\gamma p_n}=e^{\,i\gamma (N-1)/2}\sum_{n=0}^{N-1}e^{-i\gamma n},
\end{equation*}
and the remaining geometric series gives
\begin{equation*}
\sum_{n=0}^{N-1}e^{-i\gamma p_n}=\frac{\sin(N\gamma/2)}{\sin(\gamma/2)}.\label{eq:geomsum}
\end{equation*}
Hence the ideal reconstructed intensity becomes
\begin{equation}
I_{\mathrm{TRY,ideal}}^{(N)}(x_s)=I_0\left[\frac{\sin(N\gamma/2)}{\sin(\gamma/2)}\right]^2.
\label{eq:IN_ideal}
\end{equation}
Equation~(\ref{eq:IN_ideal}) is the ideal TRY grating law. Algebraically, it is identical to the textbook $N$-slit array factor, but its physical meaning is different: it describes reconstructed peaks in source space rather than outgoing diffraction orders in angle space.

The principal maxima satisfy
\begin{equation}
\gamma=2\pi m,\quad m\in\mathbb{Z},\label{eq:principal}
\end{equation}
which, using Eq.~(\ref{eq:gamma}), gives the reconstructed source-space peak positions
\begin{equation}
x_{s,m}=\frac{m\lambda z_1}{d}-\frac{z_1}{z_2}x_D.\label{eq:xsm}
\end{equation}
For a centered detector, $x_D=0$, so
\begin{equation}
x_{s,m}=\frac{m\lambda z_1}{d}.\label{eq:xsm_centered}
\end{equation}
Equivalently, using Eq.~(\ref{eq:gamma_angles}),
\begin{equation}
d(\theta_s+\theta_D)=m\lambda,\label{eq:angleorder}
\end{equation}
and for $\theta_D=0$,
\begin{equation}
d\,\theta_s=m\lambda.\label{eq:sourceorder}
\end{equation}
This is the TRY analogue of the usual grating-order law, except that it labels source-space peaks rather than outgoing angular beams. The practical consequence is that grating-like discrimination is transferred from the output side to the source side while the detector remains fixed.

The peak intensity is
\begin{equation}
I_{\max}=N^2 I_0,\label{eq:Imax}
\end{equation}
because all $N$ slit amplitudes superpose coherently at a principal maximum. The zeros satisfy
\begin{equation}
\gamma=\frac{2\pi\ell}{N},\quad\ell\not\equiv0\pmod N,\label{eq:zeros}
\end{equation}
so there are $N-1$ zeros between adjacent principal maxima. The zero-to-zero width of a principal peak in source space is
\begin{equation}
\Delta x_{s,\mathrm{zz}}=\frac{2\lambda z_1}{Nd},\label{eq:width}
\end{equation}
showing that increasing $N$ sharpens the reconstructed source-space peaks in the same way that increasing $N$ sharpens classical grating orders. In the TRY setting, however, this narrowing corresponds to improved source discrimination rather than improved output-angle dispersion. 

\begin{figure}[htbp]
\includegraphics[width=0.86\columnwidth]{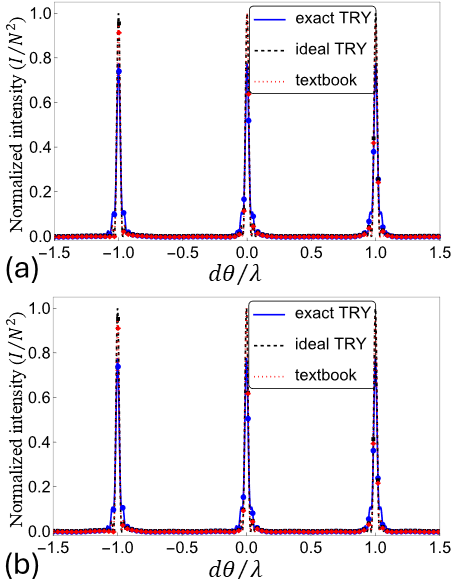}
\caption{Comparison of the exact TRY (blue solid), the ideal TRY grating law (black dashed), and the textbook Fraunhofer result with single-slit envelope (red dotted) for (a) $N=30$ and (b) $N=31$, using the same normalized coordinate and parameter conventions as in Fig.~\ref{fig:comp_smallN}.}
\label{fig:comp_largeN}
\end{figure}

Figures~2(b) and 3 illustrate how this large-\(N\) behavior emerges in practice. For \(N=4\), the exact TRY response already departs visibly from the ideal source-space grating law, while the textbook curve is distinguished mainly by envelope suppression. For \(N=30\) and \(N=31\), the principal peaks become much narrower, so the source-space selectivity predicted by Eq.~\eqref{eq:width} is immediately visible. At the same time, the exact TRY curves remain measurably distinct from the ideal array factor because the quadratic phase continues to redistribute weight among the principal peaks and nearby sidelobes. The large-$N$ regime therefore makes the central tradeoff of multi-slit TRY especially clear: increasing \(N\) sharpens source-space resolution, but also enhances sensitivity to aperture-wide quadratic-phase structure.

This ideal-limit structure also suggests direct applications. Because different source positions or source-side opening angles map to distinct reconstructed peaks, the TRY array naturally supports source-space multiplexing and source-side channel discrimination. The same fixed-port architecture is also attractive for compact interferometric processors in integrated or miniaturized platforms, since multipath discrimination can be converted into a single-detector readout \cite{SmitVanDam1996,Miller2013,Miller2020}. Finally, Eq.~(\ref{eq:width}) shows that large compensated arrays offer higher source-space resolution~\cite{BornWolf,LoewenPopov,Popov2012}, although that advantage is accompanied by increased sensitivity to the quadratic phase outside the compensated regime~\cite{LoewenPopov,Chanan1998,WenDiff2026}.

For $N=3$, Eq.~(\ref{eq:IN_ideal}) reduces to
\begin{equation}
I_{\mathrm{TRY,ideal}}^{(3)}(x_s)=I_0\left(1+2\cos\gamma\right)^2,\label{eq:I3ideal}
\end{equation}
which is precisely the ideal three-slit law recovered from Eq.~(\ref{eq:I3}) when the quadratic phase has been neutralized.

\section{Talbot-like revivals for periodic TRY arrays}
The finite-array results above already show that the essential beyond-two-slit physics is carried by the quadratic Fresnel phase $\beta$. For a strictly periodic array, the same quadratic structure produces a qualitatively new effect: Talbot-like revivals in the reconstructed TRY response. This phenomenon is closely related to the conventional Talbot effect in its mathematical origin, but it differs in both geometry and observable. In classical Talbot optics~\cite{Patorski1989,BerryKlein1996,WenTalbot2013,Testorf2010,Mansuripur2011}, the revived object is a propagated transverse field or intensity profile of a periodic grating. In the present TRY setting, by contrast, the primary observable remains a source-labeled fixed-detector response.

\paragraph*{Periodic-array limit.}
Consider an infinite periodic slit array with slit coordinates
\begin{equation}
x_n=nd,\quad n\in\mathbb{Z},\label{eq:infcoords}
\end{equation}
where $d$ is the array period. Under the same narrow-slit and paraxial assumptions used in the previous sections, the corresponding TRY field amplitude is
\begin{equation}
E_\infty(x_s;x_D)=Ae^{i\Phi_c}\sum_{n=-\infty}^{\infty}e^{i\beta n^2-i\gamma n},\label{eq:Einf}
\end{equation}
where $A$ is the slit amplitude, $\Phi_c$ is the common phase, and $\gamma$ and $\beta$ are the linear and quadratic phase parameters defined in Sec.~II. The reconstructed response is therefore
\begin{equation}
I_{\mathrm{TRY}}^{(\infty)}(x_s)\propto\left|\sum_{n=-\infty}^{\infty}e^{i\beta n^2-i\gamma n}
\right|^2.\label{eq:Iinf}
\end{equation}
For a strictly infinite array, the sum is understood in the usual distributional sense, or equivalently as the limit of a weak apodization that is removed at the end. The important point is that the periodic TRY problem is governed by a lattice sum with a discrete quadratic phase, which is precisely the mathematical structure underlying Talbot revivals.

\paragraph*{Full revival in the strict fixed-detector protocol.}
The first special case occurs when
\begin{equation}
\beta=2\pi\ell,\quad\ell\in\mathbb{Z}.\label{eq:fullrevivalbeta}
\end{equation}
Then $e^{i\beta n^2}=1$ for every integer $n$, and Eq.~(\ref{eq:Einf}) reduces to
\begin{align}
E_\infty(x_s;x_D)&=Ae^{i\Phi_c}\sum_{n=-\infty}^{\infty}e^{-i\gamma n}\nonumber\\
&=2\pi Ae^{i\Phi_c}\sum_{m=-\infty}^{\infty}\delta(\gamma-2\pi m),\label{eq:fullcomb}
\end{align}
where the last step follows from the Poisson-comb identity. Thus the infinite periodic TRY array revives into a periodic comb in the source-controlled phase $\gamma$.

Using Eq.~\eqref{eq:beta} together with $k=2\pi/\lambda$, Eq.~(\ref{eq:fullrevivalbeta}) becomes
\begin{equation}
\frac{1}{z_1}+\frac{1}{z_2}=\frac{2\ell\lambda}{d^2}.\label{eq:fullrevivalgeom}
\end{equation}
This unique reciprocal-distance law is the periodic TRY full-revival condition. For \(\ell\neq 0\), Eq.~\eqref{eq:fullrevivalgeom} may be written in the Gaussian thin-lens form
\begin{equation}
\frac{1}{z_1}+\frac{1}{z_2}=\frac{1}{f_\ell},\quad f_\ell=\frac{d^2}{2\ell\lambda},\label{eq:thinlens}
\end{equation}
with $f_\ell$ an effective discrete focal length set by the periodic array. The case $\ell=0$ corresponds to the trivial zero-quadratic-phase limit. This lens-like rewriting is geometrically useful, but in the strict TRY protocol the primary observable remains the source-space comb implied by Eq.~(\ref{eq:fullcomb}). At the same time, Eq.~\eqref{eq:thinlens} suggests possible order-resolved diffractive imaging~\cite{Valley2010}, wavelength-selective focusing~\cite{Li2022}, autofocus, geometry calibration, and single-pixel implementations in coherent bands such as THz~\cite{Chan2008} and microwave~\cite{Hunt2013}, where fixed-detector architectures are especially natural.

For a centered detector ($x_D=0$), one has
\begin{equation}
\gamma=kd\,\frac{x_s}{z_1},
\end{equation}
so the full-revival peaks satisfy
\begin{equation}
kd\,\frac{x_s}{z_1}=2\pi m,\quad m\in\mathbb{Z},
\end{equation}
or equivalently
\begin{equation}
x_s^{(m)}=\frac{m\lambda z_1}{d}.\label{eq:fixeddetectorcomb}
\end{equation}
Thus the directly reconstructed full-revival signal is a periodic comb in source space. This is the cleanest periodic-array analogue of the ideal source-space grating law derived earlier for finite $N$. Figure~4 shows this explicitly: the full-revival curve is not a transverse self-image, but a source-space peak family whose spacing is set by \(\lambda z_1/d\).

\paragraph*{Half revival and shifted source-space comb.}
A second special case occurs when
\begin{equation}
\beta=\pi(2\ell+1).\label{eq:halfrevivalbeta}
\end{equation}
Since $n^2$ and $n$ have the same parity for integer $n$, one has $e^{i\pi n^2}=e^{i\pi n}=(-1)^n$, and therefore
\begin{align}
E_\infty(x_s;x_D)&=Ae^{i\Phi_c}\sum_{n=-\infty}^{\infty}e^{-i(\gamma-\pi)n}\nonumber\\
&=2\pi Ae^{i\Phi_c}\sum_{m=-\infty}^{\infty}\delta\!\left[\gamma-(2m+1)\pi\right].\label{eq:halfcomb}
\end{align}
Compared with Eq.~\eqref{eq:fullcomb}, Eq.~\eqref{eq:halfcomb} tells that the comb is shifted by half a period in $\gamma$. For a centered detector, this gives the shifted source-space peaks
\begin{equation}
x_s^{(m+\frac12)}=\left(m+\frac12\right)\frac{\lambda z_1}{d}.\label{eq:halfcombxs}
\end{equation}
This is the direct source-space analogue of the half-period shift familiar from the classical Talbot effect. Figure~4 makes this point intuitive: the half-revival pattern is not a new envelope, but the same source-space comb displaced by half a period.

\paragraph*{Fractional revivals.}
More generally, let
\begin{equation}
\beta=\frac{2\pi p}{q},\quad p,q\in\mathbb{Z},\quad\gcd(p,q)=1.\label{eq:rationalbeta}
\end{equation}
Then Eq.~(\ref{eq:Einf}) becomes
\begin{equation}
E_\infty(x_s;x_D)=Ae^{i\Phi_c}\sum_{n=-\infty}^{\infty}\exp\!\left(i\frac{2\pi p}{q}n^2-i\gamma n\right).\label{eq:Efracstart}
\end{equation}
Now write
\[
n=lq+r,\,r=0,1,\dots,q-1,\,l\in\mathbb{Z}.
\]
Substituting this into Eq.~(\ref{eq:Efracstart}) gives
\begin{equation}
E_\infty(x_s;x_D)=Ae^{i\Phi_c}\sum_{r=0}^{q-1}e^{\,i2\pi p r^2/q-i\gamma r}
\sum_{l=-\infty}^{\infty}e^{-i\gamma ql}.\label{eq:Efracseparated}
\end{equation}
To perform the sum over $l$, we again use the Poisson-comb identity,
\begin{equation*}
\sum_{l=-\infty}^{\infty}e^{-i\gamma ql}=\frac{2\pi}{q}\sum_{m=-\infty}^{\infty}
\delta\!\left(\gamma-\frac{2\pi m}{q}\right).
\end{equation*}
Substituting this into Eq.~(\ref{eq:Efracseparated}) yields
\begin{eqnarray}
E_\infty(x_s;x_D)=\frac{2\pi A e^{i\Phi_c}}{q}\sum_{m=-\infty}^{\infty}G_m^{(p,q)}
\delta\!\left(\gamma-\frac{2\pi m}{q}\right),\label{eq:fractionalcomb}
\end{eqnarray}
where
\begin{equation}
G_m^{(p,q)}=\sum_{r=0}^{q-1}\exp\!\left[i\frac{2\pi pr^2}{q}-i\frac{2\pi mr}{q}\right]
\label{eq:gausssum}
\end{equation}
is a quadratic Gauss sum. Equation~(\ref{eq:fractionalcomb}) is the source-space analogue of the fractional Talbot effect: rational values of $\beta/2\pi$ generate a finer revival comb whose weights are set by quadratic Gauss phases. As detailed in the Appendix, the sum over $l$ is what creates the comb structure, while the remaining finite sum over \(r\) supplies the Gauss-sum weights.

For a centered detector, the corresponding fractional source-space peaks satisfy
\begin{equation}
kd\,\frac{x_s}{z_1}=\frac{2\pi m}{q},
\end{equation}
so
\begin{equation}
x_s^{(m,q)}=m\,\frac{\lambda z_1}{q\,d}.\label{eq:fractionalxs}
\end{equation}
Thus the periodic TRY array supports a hierarchy of full and fractional revivals directly in source space.

\begin{figure}[htbp]
\includegraphics[width=0.86\columnwidth]{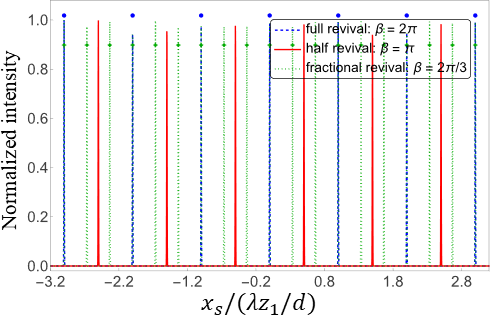}
\caption{Periodic-array TRY revivals in the strict fixed-detector setting $(x_D=0)$: full revival ($\beta=2\pi$, blue dashed with circles), half revival ($\beta=\pi$, red solid), and fractional revival ($\beta=2\pi/3$, green dotted with diamonds). The horizontal axis is the normalized source coordinate $x_s/(\lambda z_1/d)$, so the peak combs occur at $x_s=m\lambda z_1/d$ for full revival, $x_s=(m+\frac{1}{2})\lambda z_1/d$ for half revival, and $x_s=m\lambda z_1/(3d)$ for the $q=3$ fractional case. The curves were generated from the Gaussian-apodized periodic TRY sum $I^{(\infty)}_{\rm TRY}\propto\left|\sum^{n_{\rm max}}_{n=-n_{\rm max}}e^{-n^2/(2\sigma^2)}e^{i\beta n^2-i\gamma n}\right|^2$ with $n_{\rm max}=120$, $\sigma=40$, $\gamma=2\pi x_s/(\lambda z_1/d)$, and each curve normalized to unit maximum over the plotted range.}
\end{figure}

Figure~4 shows this clearly for the \(q=3\) case. Relative to the full revival, the fractional revival refines the source-space comb by a factor of \(q\), reducing the peak spacing from \(\lambda z_1/d\) to \(\lambda z_1/(qd)\). This is the source-space manifestation of the finer comb structure in Eq.~\eqref{eq:fractionalcomb}. The visual point is important: the fractional TRY revival is not simply a weaker copy of the full revival, but a denser source-space peak family whose weights are controlled by the quadratic Gauss phases in Eq.~\eqref{eq:gausssum}. In this sense, the periodic TRY array supports a genuine hierarchy of revival scales, rather than only a few isolated special cases.

\paragraph*{Difference from the conventional Talbot effect.}
The similarity to classical Talbot physics is clear: in both cases, a periodic structure together with a quadratic phase produces full and fractional revival. The difference is equally important. In conventional Talbot optics, the relevant condition is the propagation-distance law $z=qz_T$, where $z_T=2d^2/\lambda$ is the Talbot length, and the revived object is a transverse field or intensity profile after propagation. In the present TRY geometry, by contrast, the full-revival condition is instead the reciprocal-distance law in Eq.~(\ref{eq:fullrevivalgeom}), involving both the source and detector arms, and the revived quantity is a source-labeled fixed-detector response. Traditional Talbot optics is therefore a lensless self-imaging effect in physical space, whereas periodic TRY revival is a hybrid-correlation revival in source space.

This distinction also clarifies the role of imaging. In the strict fixed-detector TRY protocol studied here, $x_D$ is held fixed, so the primary result is the source-space comb in Eqs.~(\ref{eq:fixeddetectorcomb}), \eqref{eq:halfcomb}, and (\ref{eq:fractionalxs}). In that setting, there is no detector-coordinate magnification to define. Only in an extended two-coordinate formulation, in which the detector position is also allowed to vary, does the same revival condition define an image-like straight-line branches in the $(x_s,x_D)$ plane and admit a magnification interpretation.

\paragraph*{Extended two-coordinate interpretation and fixed-detector slices.}
Although the strict TRY protocol keeps the detector coordinate $x_D$ fixed, the full-revival comb condition can be rewritten in a form that makes the underlying geometry more transparent. At full revival, Eqs.~(\ref{eq:fullcomb}) and (\ref{eq:fullrevivalgeom}) imply
\begin{equation}
\gamma=2\pi m,\quad m\in\mathbb{Z},
\end{equation}
so, using Eq.~\eqref{eq:gamma}, 
\begin{equation}
\frac{x_s}{z_1}+\frac{x_D}{z_2}=\frac{m\lambda}{d}.\label{eq:linecondition}
\end{equation}
Equation~(\ref{eq:linecondition}) defines a family of straight lines in the $(x_s,x_D)$ plane. Solving for the detector coordinate gives
\begin{equation}
x_D^{(m)}(x_s)=-\frac{z_2}{z_1}\,x_s+m\,\frac{\lambda z_2}{d},\label{eq:detectorbranches}
\end{equation}
where $m$ labels the branch order. The $m=0$ branch,
\begin{equation}
x_D^{(0)}(x_s)=-\frac{z_2}{z_1}\,x_s,\label{eq:zeroorderbranch}
\end{equation}
has the form of an ordinary imaging relation with slope
\begin{equation}
\frac{dx_D^{(0)}}{dx_s}=-\frac{z_2}{z_1}.\label{eq:branchslope}
\end{equation}
Thus, in this extended two-coordinate view, the periodic TRY revival supports an imaging-like branch family with effective magnification
\begin{equation}
\mathcal M=-\frac{z_2}{z_1}.\label{eq:magnification}
\end{equation}
The higher-order branches with $m\neq 0$ are translated replicas of the same linear mapping, separated at fixed $x_s$ by
\begin{equation}
\Delta x_D=\frac{\lambda z_2}{d}.\label{eq:deltaxD}
\end{equation}

The strict fixed-detector TRY protocol is recovered by taking a horizontal slice through this branch family. Solving Eq.~(\ref{eq:linecondition}) instead for the source coordinate gives
\begin{equation}
x_s^{(m)}(x_D)=-\frac{z_1}{z_2}\,x_D+m\,\frac{\lambda z_1}{d}.\label{eq:sourceslices}
\end{equation}
For a centered detector ($x_D=0$), this reduces to the source-space comb already given in Eq.~(\ref{eq:fixeddetectorcomb}), with spacing
\begin{equation}
\Delta x_s=\frac{\lambda z_1}{d}.\label{eq:deltaxs}
\end{equation}
More generally, changing the detector position does not destroy the comb; it simply shifts the comb origin linearly. The shift rate is
\begin{equation}
\frac{\partial x_s^{(m)}}{\partial x_D}=-\frac{z_1}{z_2},\label{eq:combshift}
\end{equation}
which is the inverse slope of Eq.~(\ref{eq:branchslope}). Thus, even within a fixed-detector framework, a family of measurements taken at several detector positions could reconstruct the full branch geometry quantitatively.

The same structure persists at fractional revival. When $\beta=2\pi p/q$, the comb condition becomes
\begin{equation}
\gamma=\frac{2\pi m}{q},
\end{equation}
so the corresponding branch family is
\begin{equation}
x_D^{(m,q)}(x_s)=-\frac{z_2}{z_1}\,x_s+m\,\frac{\lambda z_2}{q\,d},\label{eq:fractionaldetectorbranches}
\end{equation}
and the fixed-detector slices are
\begin{equation}
x_s^{(m,q)}(x_D)=-\frac{z_1}{z_2}\,x_D+m\,\frac{\lambda z_1}{q\,d}.\label{eq:fractionalsourceslices}
\end{equation}
Thus the fractional Talbot-like revival refines both the branch spacing and the source-space comb spacing by a factor of $q$.

\begin{figure}[htbp]
\includegraphics[width=0.86\columnwidth]{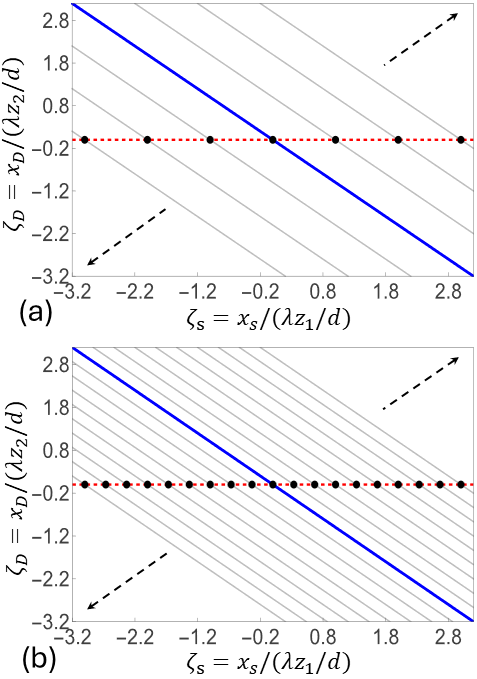}
\caption{Branch geometry for the periodic TRY revival in normalized coordinates. Panel (a) shows the full-revival branches $\zeta_D=-\zeta_s+m$ and panel (b) shows the fractional $q=3$ branches $\zeta_D=-\zeta_s+m/3$, with $\zeta_s=x_s/(\lambda z_1/d)$ and $\zeta_D=x_D/(\lambda z_2/d)$. In each panel, the gray lines are the branch family, the blue line is the central branch, the red dashed line is the fixed-detector slice $\zeta_D=0$, and the black dots are the resulting source-space revival peaks at the intersections. The plots use the window $-3.2\leq\zeta_s,\zeta_D\leq3.2$; panel (a) includes branches $m=-3,\cdots,3$, while panel (b) includes $m=-9,\cdots,9$, corresponding to offsets $m/3$.}
\end{figure}

The geometric meaning of this construction is summarized in Fig.~5. The gray oblique lines represent the branch families defined by Eqs.~\eqref{eq:detectorbranches} and \eqref{eq:fractionaldetectorbranches}, the red dashed line is the strict fixed-detector slice \(x_D=0\), and the black intersection points are exactly the source-space revival peaks measured in the strict TRY protocol. The figure makes one conceptual point particularly clear: the directly observed source-space comb is a one-dimensional slice of a larger two-coordinate revival geometry. In the full-revival case, the slice produces the integer comb of Eq.~\eqref{eq:fixeddetectorcomb}; in the \(q=3\) fractional case, it produces the finer comb of Eq.~\eqref{eq:fractionalxs}. This is why the imaging-like interpretation is geometrically useful, but should not be confused with the primary fixed-detector observable.

\paragraph*{Practical implications.}
The periodic TRY revivals suggest several practical possibilities. First, the reciprocal-distance law in Eq.~(\ref{eq:fullrevivalgeom}) provides a sharp operating condition for geometry calibration, autofocus, and focus locking in a fixed-detector architecture~\cite{Valley2010,Chan2008,Hunt2013}. Second, the full and fractional source-space combs in Eqs.~(\ref{eq:fixeddetectorcomb})--(\ref{eq:fractionalxs}) suggest periodic source-space routing, channel tagging, and comb-like multiplexing~\cite{SmitVanDam1996,Miller2020}. Third, the extended branch relations in Eqs.~(\ref{eq:detectorbranches}) and (\ref{eq:fractionaldetectorbranches}) indicate that a generalized two-coordinate TRY architecture could support replicated image channels and fractional image branches. Finally, because the same periodic structure supports both revival and off-revival quadratic-phase sensitivity, one can envision hybrid operating modes in which the array is used both for source-space discrimination and for phase-diagnostic metrology~\cite{Roddier1988,Lay2004,WenNull2026}.

Taken together, these results show that the periodic TRY array is not merely ``Talbot-like'' in a loose sense. Rather, it supports a genuine quadratic-phase revival hierarchy in source space, while remaining fully consistent with the fixed-detector hybrid-correlation logic of TRY. In an extended two-coordinate view, the same revival condition also encodes an imaging-like branch structure, but the directly measured quantity in the strict protocol remains the source-space revival comb.

\section{Comparison with textbook three-slit, grating, and Talbot optics}
In standard Fraunhofer optics~\cite{BornWolf,GoodmanFO,Hecht,LoewenPopov}, the textbook three-slit and $N$-slit intensities are
\begin{equation}
I_{\mathrm{cl}}^{(3)}(\theta)=I_1(\theta)\left(1+2\cos\phi\right)^2,\;\;
\phi=kd\sin\theta,\label{eq:classical3}
\end{equation}
and
\begin{equation}
I_{\mathrm{cl}}^{(N)}(\theta)=I_1(\theta)\left[\frac{\sin(N\phi/2)}{\sin(\phi/2)}\right]^2,
\;\;\phi=kd\sin\theta,\label{eq:classicalN}
\end{equation}
where $\theta$ is the observation angle, $d$ is the slit spacing, and
\begin{equation}
I_1(\theta)=I_1(0)\left[\frac{\sin\!\left(\frac{\pi a}{\lambda}\sin\theta\right)}
{\frac{\pi a}{\lambda}\sin\theta}\right]^2\label{eq:single_envelope}
\end{equation}
is the single-slit diffraction envelope for slit width $a$, with $I_1(0)$ the on-axis single-slit intensity \cite{BornWolf,GoodmanFO,Hecht,LoewenPopov}. These formulas describe first-order interference in angle space: the array factor determines the fine interference structure, while the single-slit term provides the diffraction envelope.

The multi-slit TRY laws derived in this work differ from these textbook Fraunhofer results in three essential respects. First, the independent variable is different. In classical optics, the pattern is read out in the observation angle $\theta$ or, equivalently in a screen coordinate. In TRY, the response is reconstructed in source space, through the source coordinate $x_s$ or the source-side opening angle $\theta_s$. Second, the meaning of an order is different. In a classical grating, $d\sin\theta=m\lambda$ labels real outgoing diffraction directions. In TRY, the analogous condition labels reconstructed source-space maxima rather than distinct propagating beams. Third, the leading correction away from the ideal array law is different. In classical Fraunhofer optics, the dominant modification is the single-slit envelope \(I_1(\theta)\). In the narrow-slit TRY model, the leading nontrivial modification is instead the quadratic Fresnel phase \(\beta\).

This last distinction is especially important. In the classical formulas, the multi-slit factor is multiplied by the envelope \(I_1(\theta)\) because the observable is the first-order intensity distribution on an observation plane. In the present TRY formulation, by contrast, the reconstructed quantity \(I_{\mathrm{TRY}}^{(N)}(x_s)\) is a source-labeled fixed-detector response rather than an observation-plane intensity profile. For that reason, the exact TRY law is not written as a classical single-slit envelope multiplying an array factor. Within the present narrow-slit model, the dominant departure from the ideal source-space grating law is therefore the aperture-wide quadratic phase \(\beta\), not the textbook envelope \(I_1(\theta)\).

These finite-array differences are illustrated in Figs.~2 and 3 and were discussed in the previous sections where the corresponding results were derived. For small arrays, the comparison isolates the onset of genuinely beyond-two-slit quadratic-phase physics, most clearly through the lifted three-slit minima and the deformation of the exact TRY curve away from the ideal source-space grating law. For larger arrays, the same comparison shows that sharper source-space selectivity and stronger quadratic-phase sensitivity emerge together. The essential point is that the TRY/classical contrast is not primarily a difference in peak locations, but a difference in what modifies the ideal array law: in classical Fraunhofer optics the main correction is the single-slit envelope, whereas in TRY it is the aperture-wide Fresnel phase.

The comparison with Talbot physics is equally instructive. In conventional Talbot optics, a periodic grating produces full and fractional self-images of a transverse field distribution after propagation over special distances, with Talbot length \(z_T=2d^2/\lambda\) in the standard paraxial setting. In the periodic TRY problem, the same quadratic-phase logic produces full and fractional revival, but the primary observable is different: it is the source-labeled fixed-detector response in Eq.~\eqref{eq:Iinf}, not a propagated transverse intensity. Correspondingly, the full-revival condition is the reciprocal-distance law in Eq.~\eqref{eq:fullrevivalgeom}, not the conventional propagation-distance law. Figures~4 and 5 make this distinction concrete. Figure~4 shows the directly reconstructed source-space revival combs in the strict fixed-detector protocol, while Fig.~5 shows how those combs arise as slices of the larger branch geometry in the extended two-coordinate interpretation.

This also clarifies the role of imaging. In conventional Talbot optics, image replication is part of the primary physical picture. In the strict TRY protocol, however, the directly reconstructed observable is the source-space comb described by Eqs.~(\ref{eq:fixeddetectorcomb}), (\ref{eq:halfcombxs}), and (\ref{eq:fractionalxs}). The imaging-like branch structure discussed in Sec.~V appears only in an extended two-coordinate interpretation in which the detector coordinate is also allowed to vary. The main fixed-detector result is therefore a source-space revival hierarchy, not a transverse self-image.

The practical contrast is therefore sharp. A classical grating is naturally an output-angle disperser, and a classical Talbot grating is naturally a near-field self-imaging system. A multi-slit TRY array, by contrast, is naturally a fixed-detector source-space analyzer. In the compensated regime, it provides grating-like source discrimination without requiring angle-resolved detection. Outside that regime, the same geometry becomes intrinsically sensitive to quadratic phase across the aperture. In the periodic limit, this sensitivity organizes into a Talbot-like revival hierarchy in source space. That combination of source-space grating behavior, quadratic-phase sensitivity, and periodic-array revival is what most clearly distinguishes the multi-slit TRY system from textbook three-slit, grating, and Talbot optics.

\section{Conclusion}
In summary, here we have developed a compact theory of time-reversed Young interference beyond the symmetric two-slit geometry. The central result is that, for three or more slits, the reconstructed fixed-detector response is governed not only by the linear array phase but also by a quadratic Fresnel phase that survives across the aperture and reshapes the source-space interference law. In the compensated regime, the multi-slit TRY response reduces to an ideal grating-like law in source space, so the array naturally functions as a fixed-detector analyzer of source position or source-side opening angle. Outside that regime, the surviving quadratic phase makes the response directly sensitive to defocus, wavefront curvature, and aperture-scale phase nonuniformity.

In the periodic-array limit, the same quadratic phase generates full and fractional Talbot-like revivals in source space. These revivals share the quadratic-phase origin as the conventional Talbot effect, but differ in both geometry and observable: the primary result is a hybrid-correlation revival of a source-labeled fixed-detector response, rather than a transverse-field self-image. The overall picture is therefore clear. A classical grating disperses light in output-angle space, and a classical Talbot grating self-images in the near field, whereas a multi-slit TRY array reconstructs interference in source space while simultaneously revealing aperture-wide quadratic-phase structure. This combination is what gives the multi-slit TRY geometry its distinctive physical character and practical potential.

\section*{Acknowledgments}
This work was partially supported by startup funds from Binghamton University and NSF ExpandQISE 2329027.

\appendix

\section*{Detailed derivation of Eqs.~\eqref{eq:fractionalcomb} and \eqref{eq:gausssum}}
As shown in Sec.~V, after substituting $n=lq+r$ into Eq.~(\ref{eq:Efracstart}) one obtains
\begin{equation*}
E_\infty(x_s;x_D)=Ae^{i\Phi_c}\sum_{r=0}^{q-1}\sum_{l=-\infty}^{\infty}\exp\!\left[
i\frac{2\pi p}{q}(lq+r)^2-i\gamma(lq+r)\right].
\end{equation*}
Expanding \((lq+r)^2=l^2q^2+2lqr+r^2\), the exponent becomes
\[
i2\pi p\,l^2 q+i4\pi p\,lr+i\frac{2\pi p r^2}{q}-i\gamma ql-i\gamma r.
\]
Since $p,q,l$, and $r$ are integers, the first two terms are integer multiples of \(2\pi i\), and their exponentials are therefore unity. The field amplitude then reduces to
\begin{equation}
E_\infty(x_s;x_D)=Ae^{i\Phi_c}\sum_{r=0}^{q-1}e^{\,i2\pi pr^2/q-i\gamma r}\sum_{l=-\infty}^{\infty}e^{-i\gamma ql}.\label{eq:Efracseparated}
\end{equation}
At this stage, the role of the index \(l\) is explicit: it generates the comb structure, while the finite sum over $r$ will produce the Gauss-sum weights.

We now evaluate the $l$-sum using the Poisson-comb identity
\begin{equation*}
\sum_{l=-\infty}^{\infty}e^{-i\alpha l}=2\pi\sum_{m=-\infty}^{\infty}\delta\!\left(\alpha-2\pi m\right).
\end{equation*}
With $\alpha=\gamma q$, this gives
\[
\sum_{l=-\infty}^{\infty}e^{-i\gamma ql}=2\pi\sum_{m=-\infty}^{\infty}\delta\!\left(\gamma q-2\pi m\right).
\]
Next, use the scaling property of the Dirac delta,
\[
\delta(ax)=\frac{1}{|a|}\delta(x),
\]
with $a=q$, to write
\[
\delta\!\left(\gamma q-2\pi m\right)=\frac{1}{q}\delta\!\left(\gamma-\frac{2\pi m}{q}\right).
\]
Therefore,
\[
\sum_{l=-\infty}^{\infty}e^{-i\gamma ql}=\frac{2\pi}{q}\sum_{m=-\infty}^{\infty}\delta\!\left(\gamma-\frac{2\pi m}{q}\right).
\]
Substituting this result back into Eq.~(\ref{eq:Efracseparated}) yields
\[
E_{\infty}(x_s;x_D)=Ae^{i\Phi_c}\sum^{q-1}_{r=0}e^{i\frac{2\pi pr^2}{q}-i\gamma r}\left[\frac{2\pi}{q}\sum^{\infty}_{m=-\infty}\delta\!\left(\gamma-\frac{2\pi m}{q}\right)\right].
\]
We now use the standard distribution identity
\[
f(\gamma)\delta(\gamma-\gamma_m)=f(\gamma_m)\delta(\gamma-\gamma_m),\;\gamma_m=\frac{2\pi m}{q},
\]
which allows the factor $e^{-i\gamma r}$ to be evaluated at the support of each delta function:
\[
e^{-i\gamma r}\delta\!\left(\gamma-\frac{2\pi m}{q}\right)=e^{-i(2\pi m/q)r}\delta\!\left(\gamma-\frac{2\pi m}{q}\right).
\]
Then by some algebraic manipulation, one readily arrives at Eqs.~\eqref{eq:fractionalcomb} and \eqref{eq:gausssum} given in Sec.~V.

\end{document}